\documentclass[referee]{aa}
\usepackage{graphicx}
\usepackage{natbib}
\bibpunct{(}{)}{;}{a}{}{,}

\begin{document}

\hyphenation{Fe-bru-ary Gra-na-da mo-le-cu-le mo-le-cu-les}
\title{Sulphur chemistry and molecular shocks: the case of NGC1333-IRAS2}
%     \title{Sulphur emission along the outflow of IRAS2-NGC1333}

\subtitle{}

\author{V. Wakelam\inst{1}, C. Ceccarelli\inst{2}, A. Castets\inst{1}, B. Lefloch\inst{2},
L. Loinard\inst{3}, A. Faure\inst{2},
N. Schneider\inst{1}, J-J. Benayoun\inst{2}
}
\institute{
L3AB Observatoire de Bordeaux, BP 89, 33270 Floirac, France
\and 
Laboratoire d'Astrophysique, Observatoire de Grenoble - BP 53,
F-38041 Grenoble cedex 09, France
\and 
Centro de Radioastronom\'ia y Astrof\'isica, 
Universidad Nacional Aut\'onoma de M\'exico, Apartado Postal Postal 
72-3 (Xangari), 58089 Morelia, 
Michoac\'an, Mexico
}

\offprints{wakelam@mps.ohio-state.edu}
\date{Received 17 December 2004 / Accepted 23 February 2005 }
\abstract{
We present SO and SO$_2$ observations in the region of the low mass 
protostar IRAS2/NGC1333. The East-West outflow originating from this 
source has been mapped in four transitions of both SO and SO$_2$. In 
addition, CS observations published in the literature have 
been used. We compute the SO, SO$_2$ and CS column densities and the 
physical conditions at several positions of the outflow using LTE and a 
non-LTE LVG approximations. The SO$_2$/SO and CS/SO abundance ratios are 
compared with the theoretical predictions of a chemical model adapted to 
the physical conditions in the IRAS2 outflow. 
     
The SO$_2$/SO abundance ratios are constant in the two lobes of the outflow 
whereas CS/SO is up to 6 times lower in the shocked gas of the East lobe 
than in the West one. The comparison with the chemical model allows us to 
constrain the age of the outflow produced by IRAS2 to $\le 5\times 
10^3$~yr. We find low densities and temperatures for the outflow of IRAS2 
($< 10^6$~cm$^{-3}$ and $\le 70$~K) from SO and SO$_2$ emission probably 
because the two molecules trace the cooled entrained material. The East lobe of the outflow shows denser gas compared to  
the West lobe. We also discuss some constraints on the depleted form of 
sulphur.
}

\titlerunning{Sulphur chemistry and molecular shocks}
\authorrunning{Wakelam et al.}
\maketitle

%________________________________________________________________

\section{Introduction}

Accretion and ejection are two apparently antithetic phenomena occurring
simultaneously during the first stages of star formation.
Both are vital in the overall process: the former for the central
object to grow, the latter to eliminate the excess angular momentum, and to allow
accretion to proceed.
In the very first phases of star formation, the ejection occurs through
collimated, fast and extended flows of material, often probed
by molecular rotational lines, and therefore refered to as molecular 
outflows and/or jets.
Indeed, the forming star being still embedded in the parental molecular cloud,
the outflowing gas strikes the ambient material, 
often causing molecular shocks at the interaction interface.
These shocks can be and indeed often are so violent that the grain
mantles are partially destroyed, so that some heavy elements, usually
frozen onto the grain surfaces, are released into the gas phase.
Depending on the shock velocity, magnetic field and pre-shock density,
the shocks may dissociate the existing molecules 
\citep[J-type shocks:][]{1989ApJ...342..306H}, or, on the contrary, form 
many molecules in the ``gently'' shocked gas
\citep[C-type shocks:][]{1980ApJ...241.1021D,2003MNRAS.343..390F}, or in the 
post-shocked gas.
The SiO molecule represents a representative case in this respect 
\citep{1997A&A...321..293S},
but several other molecules have been observed to have abundances
much greater in molecular shocks than in molecular clouds 
\citep[e.g.][]{1996ARA&A..34..111B}. The time scale of the
  ejection process compared to the protostar life is not well
  known. To date the outflows there are two possible methods. One is
  dynamical and uses the ejected matter velocity (based on the line
  profiles) and the distance from the protostar. 
  An other way to constrain the age of shocks is to study
  their chemical evolution through chemical clocks. Chemical clocks
  are abundant species whose abundance depends more on
  time compared with the dynamical evolution of the source. 
  Thus comparing the observed abundances of these species
  with their theoretical evolution should indicate the age of the
  shocked region.

Sulphur-bearing species are of particular interest, for the following
two reasons:\\
a) Sulphur is known to be severely depleted in molecular clouds, where its
measured abundance is 1000 times less than the cosmic abundance
\citep{1994A&A...289..579T}. Note that the sulphur depletion likely occurs in
the molecular cloud phase because in the diffuse medium the total abundance of
sulphur in the gas phase is close to the cosmic abundance of S. On the
other hand, SO, SO$_2$ and H$_2$S are often abundant in molecular shocks 
\citep{1997ApJ...487L..93B,1999A&A...350..659C,2004A&A...413..609W}; therefore
S-bearing molecules can be potentially good shock tracers; \\
b) Sulphur chemistry is relatively fast, with typical timescales of
some $10^4$ yr \citep{2004A&A...422..159W}; therefore, the relative abundances 
of ``appropriate''
S-bearing molecules could in principle be used to date the outflows 
\citep{2001A&A...372..899B}.\\
These two properties have long been recognized and several theoretical studies
have focused on the use of sulphur-bearing species as  ``chemical clocks'' 
\citep{1997ApJ...481..396C,1998A&A...338..713H,
2001A&A...372..899B,2003A&A...412..133V,2004A&A...413..609W,2004A&A...422..159W}.
Several observational studies have targeted the ``hot cores'' of protostars,
regions where the dust mantles evaporate because of the elevated dust
temperature. 
Unfortunately, \citet{2004A&A...422..159W} have demonstrated that the use of
S-bearing molecules as chemical clocks is far from easy. 
The main difficulty is that, indeed, the time evolution of the
S-bearing species abundances critically depends on the initial form
of the sulphur on the grain mantles, and on the exact physical conditions
of the gas.
As a result, it is very difficult to estimate the age of sources belonging 
to different 
molecular 
clouds since the initial sulphur composition could be very different and
mask any evolutionary effect on the S-bearing species abundances.
It is, however, still possible that the abundance of S-bearing molecules
in shocked gas  belonging to the same system (outflow source and
molecular cloud) can be used to date the different shock sites.
At least, one can expect that this would be easier to interpret than in 
the case of hot 
cores of different sources.
In other words, a system where ejection events occurred at different times
and caused successive shock sites could be a good benchmark for a
time-dependent sulphur chemistry study. 
For this reason we decided to study an outflow system where multiple ejection 
events are suspected to have occurred.
The basic question we want to answer is: 
{\it can the (relative) abundance of sulphur-bearing molecules identify shocks
that occurred at different times? }
The article is organized as follows.
We describe the selected outflow system (\S 2), the observations 
(\S 3), the obtained maps (\S 4) of the most abundant S-bearing molecules 
in the region and the derived relative abundances along the outflow
(\S 5). In \S 6 we describe the chemical model \citep{2004A&A...422..159W} 
applied to the outflow system studied. Finally, in \S 7, we discuss the 
results and compare the observed abundances with the predictions of the 
chemical model.

\section{NGC1333-IRAS2: source background}

The selected source, IRAS2, is situated in the NGC1333 molecular cloud,
an  active star forming region at 220 pc \citep{1990Ap&SS.166..315C}, with 
several low- and intermediate-mass protostars 
\citep{1994A&AS..106..165A,1997ApJ...488..286L}.
Evidence of the interaction between the protostars and the parental cloud (like
H$_2$ jets, Herbig-Haro objects and molecular outflows)  have been
widely reported \citep{1990A&A...239..276B,1994A&A...285L...1S,
1995ApJ...441..689B,1996A&A...306..935W,1996ApJ...471L.111L,
1996MNRAS.281L..53W,1998A&A...334..269L,1998ApJ...504L.109L,
1998A&A...335..266B,2000A&A...361..671K}.
NGC1333 hosts a few very young Class 0 sources only detectable
in the millimeter to FIR wavelength range \citep{1987MNRAS.226..461J}.

Among them, IRAS2 (IRAS 03258+3104 source of the IRAS Point Source Catalogue) 
is a low mass 
Class 0 protostar located at the edge of a cavity 
\citep{1996ApJ...471L.111L}. IRAS2 is 
indeed composed of three continuum components: A, B and C. The B and C 
components 
(IRAS2B and IRAS2C) are respectivelly at 30$''$ South-East and 30$''$ 
North-West of 
the A component (IRAS2A) \citep{2000ApJ...529..477L,2000A&A...361..671K}. 
In practice, 
IRAS2A is the most studied source among the three, and it is often refered to
as IRAS2 
\citep{1994A&A...285L...1S}. In the following, we will follow this tradition 
of calling 
IRAS2 the A component. \\
Two bipolar outflows in the North-South and East-West directions emanate 
from IRAS2, 
suggesting that it is a non-resolved binary system 
\citep{1994A&A...285L...1S}. 
The North-South outflow was first detected by \citet{1988A&A...192..153L} 
in the CO 1 - 0 transition. Later on, \citet{1994A&A...285L...1S} showed that 
this outflow 
is not very collimated \citep[see also][ for a detailled study of this 
outflow]{2000A&A...361..671K}. These authors also discovered another outflow 
coming out 
from IRAS2 and oriented East-West. This outflow is very powerful, collimated 
and has a 
high inclination in the plane of the sky, which makes the East and West 
lobes spatially 
well separated \citep{1998A&A...335..266B}. Using interferometric maps 
of the CH$_{3}$OH 
emission, \citet{1998A&A...335..266B} have proposed that the lobes of this 
outflow are 
composed of several bullets, created in successive mass loss episodes. They 
predicted a 
difference of $\sim 2 \times 10^{3}$~yr between the farthest and nearest 
bullets. 

We obtained maps of the IRAS2 East-West outflow system in four transitions of 
SO and SO$_{2}$ respectively.
The first goal is to estimate the gas temperature and density as well as 
 the column density of SO and SO$_2$ along 
the outflow. 
We also re-computed the CS column densities along the outflow, using the
line emission maps previously published in \citet{1996ApJ...471L.111L}. The 
ultimate goal 
is to compare the SO/SO$_2$/CS ratios observed along the outflow with the 
predictions of 
the chemical model by \citet{2004A&A...422..159W}, and to estimate the age 
of the shocks along the outflow. 

\section{Observations}\label{obs}

The IRAS2 region was observed with the IRAM 30m 
telescope at Pico Veleta (near Granada, Spain) during several runs in January
and August 1998 and January 1999. We performed large-scale maps at a sampling
of $12 \arcsec$ in the following  molecular transitions:
SO ($3_{2} \rightarrow 2_{1}$, $2_{3} \rightarrow 1_{2}$, $4_{3}
\rightarrow 3_{2}$ and $6_{5} \rightarrow 5_{4}$ transitions) and
SO$_{2}$ ($3_{1,3}
\rightarrow 2_{0,2}$, $10_{1,9} \rightarrow 10_{0,10}$, $5_{1,5} \rightarrow
4_{0,4}$ and $5_{2,4} \rightarrow 4_{1,3}$ transitions). 
We covered an area of $\Delta\alpha \times \Delta\delta$ = 228$''\times$ 96$''$
in SO and 216$''\times$ 84$''$ in SO$_2$. 
The relative coordinates ($\Delta\alpha, \Delta\delta$) 
used in all the maps presented here refer to the source IRAS2
($\alpha(2000.0)$ = 3$^{\rm h}$28$^{\rm m}$55$^{\rm s}$41,
$\delta(2000.0)$ = 31$^0$14$'$35.08$''$).
Table~\ref{parameters} summarizes the relevant parameters of the
observed transitions. 

%%{\bf Complementary observations were performed with the IRAM-30m telescope 
%%in the 
%%CO~J=2-1 line at position $(72\arcsec,-12\arcsec)$, in the East bow-shock 
%%of the IRAS2 protostellar outflow, and at position 
%%($(-60\arcsec,12\arcsec)$, 
%%in the Western flow. The instrumental setup was the same as above.}

All observations were performed using the position switching mode,
with an off position at $\Delta$$\alpha$ = +70$''$ and $\Delta$$\delta$
= +190$''$ from the nominal position of IRAS2 in NGC~1333. This position is 
void of molecular emission since it corresponds to the center of the 
molecular cavity described in \citet{1996ApJ...471L.111L}.
Pointing was checked 
every 1-2 hrs on the nearby continuum source 
0333+321. The pointing accuracy was found to be better than $3\arcsec$.
An autocorrelator
split into three parts was connected to three receivers working
at 1, 2 and 3 mm respectively, allowing simultaneous observations of three 
different transitions. 
The achieved spectral resolution is $\leq$ 0.12~km~s$^{-1}$.    
The weather conditions were very good, and typical system temperatures 
were in the range of 120, 200 and 300~K  at 3, 2 and 1.3 mm respectively. 
The spectra
were calibrated with the standard chopper-wheel method and are reported
here in units of main-beam brightness temperatures. 

\begin{table*}
\caption{Parameters of the observed SO and SO$_2$ transitions: the 
frequency ($\nu$), the energy of the upper level divided by the Boltzmann 
constant (E$_{up}$/k), the telescope beam, the 
beam efficiency of the telescope (B$_{\rm eff}$) and the spectral
resolution ($\Delta \nu$).\label{parameters}}
\begin{tabular}{lccccc}
\hline
\hline
Transitions & $\nu$ & E$_{up}$/k & Beam & $\rm B_{eff}$ &
$\Delta \nu$  \\
              & (GHz) & (K) & \arcsec\  & & (kHz) \\
\hline
SO $3_{2} \rightarrow 2_{1}$ & 99.299 & 9 & 24 & 0.73 & 40  \\
SO $2_{3} \rightarrow 1_{2}$ & 109.252 & 21 & 22 & 0.68 &  40 \\
SO $4_{3} \rightarrow 3_{2}$ & 138.178 & 16 & 18 & 0.54 & 40  \\
SO $6_{5} \rightarrow 5_{4}$ & 219.949 & 35 & 11 & 0.41 & 80  \\
SO$_{2}$ $3_{1,3} \rightarrow 2_{0,2}$ & 104.029 & 8 & 24 & 0.69 & 40 \\
SO$_{2}$ $10_{1,9} \rightarrow 10_{0,10}$ & 104.239 & 55 & 24 & 0.69 & 40 \\
SO$_{2}$ $5_{1,5} \rightarrow 4_{0,4}$ & 135.696 & 16 & 19 & 0.56 & 40 \\
SO$_{2}$ $5_{2,4} \rightarrow 4_{1,3}$ & 241.615 & 24 & 11 & 0.39 & 80 \\
\hline
\end{tabular}
\end{table*}

\section{Results}\label{results}
\begin{figure}
%\centering
\includegraphics[angle=270,width=15cm]{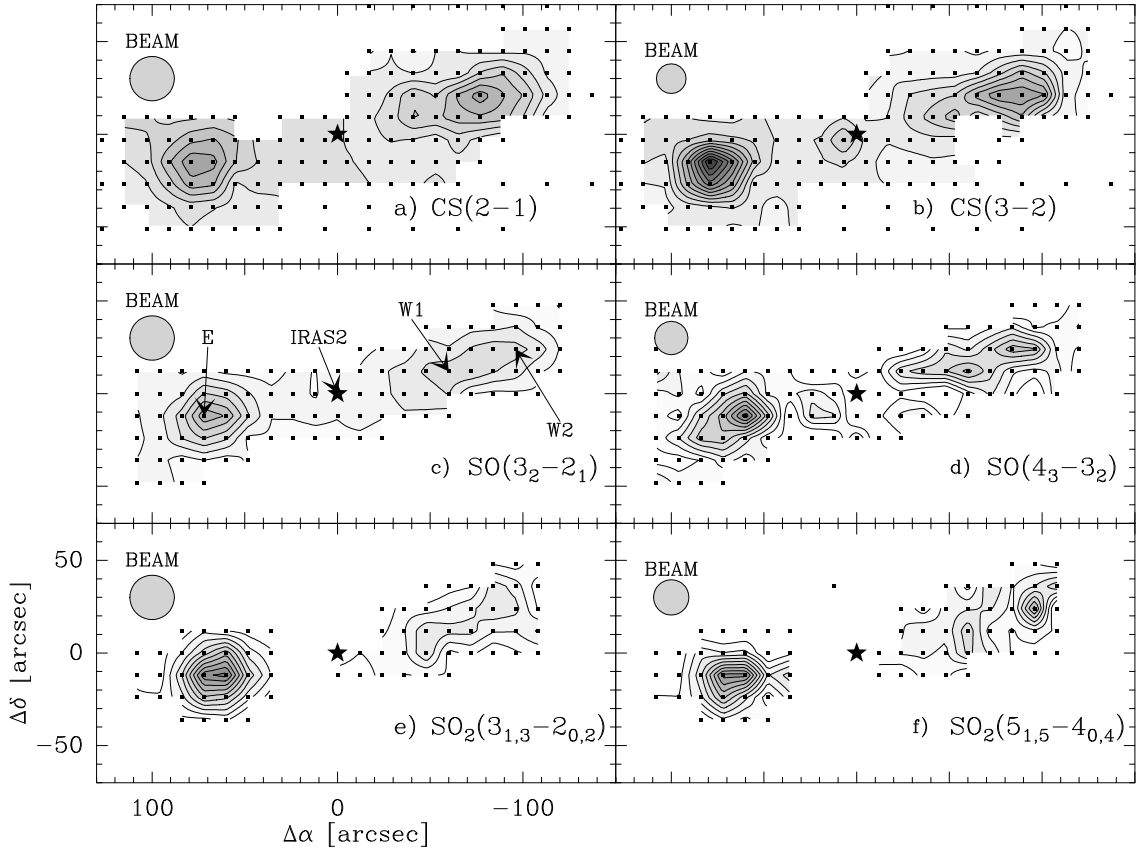}
\caption{Integrated intensity maps of the CS $2 \rightarrow 1$ (a) 
CS $3 \rightarrow 2$ (b) SO $3_{2} \rightarrow 2_{1}$ (c)
SO $4_{3} \rightarrow 3_{2}$  (d) SO$_{2}$ $3_{1,3} \rightarrow 2_{0,2}$
(e) and SO$_{2}$ $5_{1,5} \rightarrow 4_{0,4}$ (f) transitions. (a), 
(b), (c) and (d) : first level is
0.5~K~km~s$^{-1}$ with level step of 2.5~K~km~s$^{-1}$.
(e) and (f): first level is 0.2~K~km~s$^{-1}$
with level step of 0.3~K~km~s$^{-1}$. 
The CS data are taken from \citet{1996ApJ...471L.111L} whereas 
the SO and SO$_{2}$ data are from this work.
In each map, the
black points represent the observed positions. The star shows the position of
the protostellar source IRAS2. The arrows
point to the positions whose spectra are displayed in Fig.~\ref{spect}.
}
\label{4maps}
\end{figure}
\begin{figure}
\centering
\includegraphics[angle=270,width=11cm]{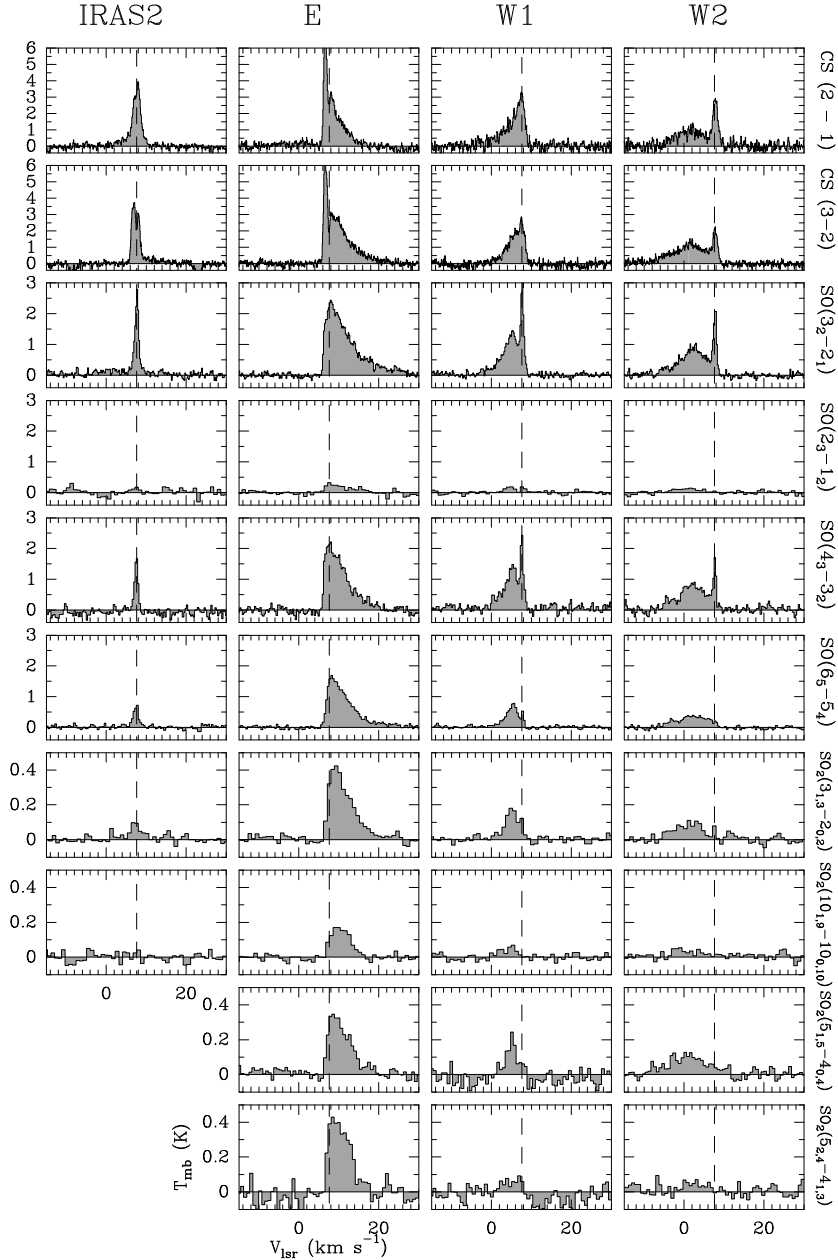}
\caption{Spectra of the SO, SO$_{2}$ (this work) and CS 
\citep{1996ApJ...471L.111L} transitions observed 
toward IRAS2 and along the outflow, in the three positions E 
($72''$, $-12''$), W1 ($-60''$, $12''$) and W2 ($-96''$, $24''$) (see text).
All shown spectra are smoothed at the largest beam $24 \arcsec$. The SO$_{2}$
$5_{1,5} \rightarrow 4_{0,4}$ and $5_{2,4} \rightarrow 4_{1,3}$ toward IRAS2 
are not represented because they cannot be smoothed (see text).
The vertical dashed line on each spectrum marks the position of the cloud
systemic velocity (v$_{LSR}$ = 7.6~km~s$^{-1}$).
}
\label{spect}
\end{figure}
\begin{figure}
\centering
\includegraphics[angle=270,width=1\linewidth]{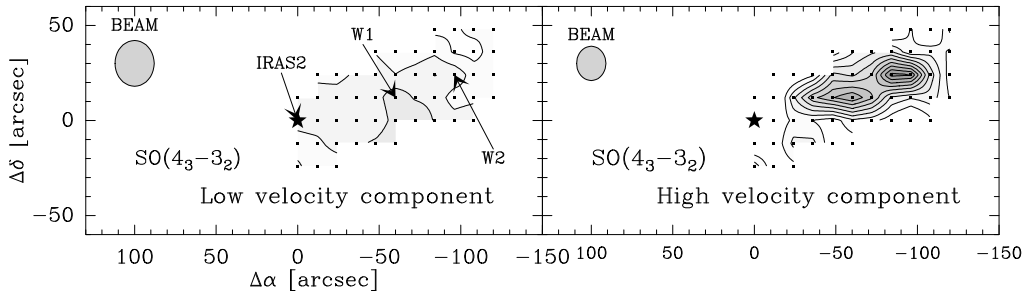}
\caption{Integrated intensity maps of the LVC and HVC 
SO $4_{3} \rightarrow 3_{2}$ transition separately traced on the left 
and right panels respectively. The first level is 0.5~km~s$^{-1}$ with 
level step of 1~km~s$^{-1}$. }
\label{solowhigh}
\end{figure}
\begin{table*}
\caption{Intensity (T$_{\rm MB}$) in K, velocity width ($\Delta v$) in 
km~s$^{-1}$ and integrated intensity ($\int T_{\rm MB} \delta v$) in 
K~km~s$^{-1}$ of the SO and SO$_2$ line profiles shown in Fig.~\ref{spect}. 
LVC and HVC mean 
the low and the high velocity component of the spectra of the west 
lobe. The linewidth of E(LVC) is fixed to 1~km~s$^{-1}$ based on the 
linewidths of the West LVC components because the LVC and HVC components 
of the East lobe are mixed. \label{intensity}}
\begin{tabular}{lcccccccc}
\hline
\hline
 &  & IRAS2 & E(LVC) & E(HVC) & W1 (LVC) & W1 (HVC) & W2 (LVC) & W2 (HVC) \\
 & $\Delta\alpha$, $\Delta\delta$ ($\arcsec$) & 0, 0 & 72, -12 & 72, -12 & -60, 
 12 & -60, 12 & -96, 24 & -96, 24 \\
\hline
SO $3_{2} \rightarrow 2_{1}$ & T$_{\rm MB}$ & 2.4$\pm$0.09 & 2.39$\pm$0.07 & 
       2.02$\pm$0.07 & 2.7$\pm$0.08 & 1.4$\pm$0.08 & 2.0$\pm$0.08 & 0.9$\pm$0.08 \\
       & $\Delta v$ & 1.1$\pm$0.03 & 1 & 5.0$\pm$0.5 & 1.1$\pm$0.09 & 
       3.2$\pm$0.2 & 1.1$\pm$0.09 & 7.2$\pm$0.2 \\
       & $\int T_{\rm MB} \delta v$ & 3.5$\pm$0.6 & 3.6$\pm$0.6 & 
       14.3$\pm$2.3 & 3.1$\pm$0.5 & 6.1$\pm$1.1 & 1.9$\pm$0.3 & 7.1$\pm$1.2 \\
SO $2_{3} \rightarrow 1_{2}$ & T$_{\rm MB}$ & $\leq$0.3 & 0.3$\pm$0.07 & 0.24$\pm$0.07 & 
       0.31$\pm$0.05 & 0.20$\pm$0.05 & $\leq$0.15 & 0.15$\pm$0.05\\
       & $\Delta v$ & - & 1 & 7.7$\pm$1.3 & 0.8$\pm$1.9 & 2.8$\pm$0.6 & - & 
       8.3$\pm$1.6 \\
       & $\int T_{\rm MB} \delta v$ & $\leq$0.6 & 0.4$\pm$0.08 & 2.0$\pm$0.7 & 
       0.3$\pm$0.1 & 0.6$\pm$0.2 & $\leq$0.3 & 1.2$\pm$0.4\\
SO $4_{3} \rightarrow 3_{2}$ & T$_{\rm MB}$ &1.8$\pm$0.1 & 2.2$\pm$0.1 & 
       1.8$\pm$0.1 & 2.6$\pm$0.2 & 1.5$\pm$0.2 & 1.7$\pm$0.2 & 0.8$\pm$0.2 \\
       & $\Delta v$ & 1.04$\pm$0.06 & 1 & 6.5$\pm$0.2 & 0.89$\pm$0.06 & 
       2.7$\pm$0.3 & 0.8$\pm$0.1 & 7.6$\pm$0.3\\
       & $\int T_{\rm MB} \delta v$ & 1.9$\pm$0.4 & 
       2.3$\pm$0.4 & 11.3$\pm$1.9 & 2.7$\pm$0.5 & 5.6$\pm$1.0 & 
       2.0$\pm$0.4 & 6.4$\pm$1.2\\
SO $6_{5} \rightarrow 5_{4}$ & T$_{\rm MB}$ & 0.69$\pm$0.08 & 
       1.7$\pm$0.1 & 1.2$\pm$0.1 & 0.56$\pm$0.05 & 0.77$\pm$0.05 & 
       0.24$\pm$0.06 & 0.38$\pm$0.06 \\
       & $\Delta v$ & 1.5$\pm$0.1 & 1 & 7.1$\pm$0.2 & 0.8$\pm$0.1 & 
       2.7$\pm$0.2 & 1 & 8.9$\pm$0.4 \\
       & $\int T_{\rm MB} \delta v$ & 1.2$\pm$0.3 & 2.4$\pm$0.3 & 8.3$\pm$1.4 & 
       0.44$\pm$0.09 & 2.6$\pm$0.5 & 0.26$\pm$0.09 & 3.6$\pm$0.7  \\
SO$_{2}$ $3_{1,3} \rightarrow 2_{0,2}$ & T$_{\rm MB}$ & 0.09$\pm$0.02 & 0.4$\pm$0.02 
       & 0.44$\pm$0.02 & 0.13$\pm$0.02 & 0.17$\pm$0.02 & 0.08$\pm$0.02 & 
       0.09$\pm$0.02 \\
       & $\Delta v$ & 2.4$\pm$0.7 & 1 & 4.6$\pm$0.4 & 1.0$\pm$0.1 & 4.3$\pm$0.3 & 1.0$\pm$0.1 & 
       9.7$\pm$1.1 \\
       & $\int T_{\rm MB} \delta v$ & 0.23$\pm$0.09 & 0.4$\pm$0.08 & 
       2.8$\pm$0.5 & 0.14$\pm$0.03 & 0.8$\pm$0.2 & 0.08$\pm$0.03 & 1.0$\pm$0.2 \\
SO$_{2}$ $10_{1,9} \rightarrow 10_{0,10}$ & T$_{\rm MB}$ & $\leq$0.06 & 0.08$\pm$0.02 & 
       0.18$\pm$0.02 & $\leq$0.05 & 0.06$\pm$0.02 & $\leq$0.6 & $\leq$0.6 \\
       & $\Delta v$ & - & 1 & 5.4$\pm$0.3 & - & 5.2$\pm$0.9 & - & - \\
       & $\int T_{\rm MB} \delta v$ & $\leq$0.2 & 0.1$\pm$0.05 & 2.8$\pm$0.5 & 
       $\leq$0.07 & 0.25$\pm$0.09 & $\leq$0.06 & $\leq$0.2 \\
SO$_{2}$ $5_{1,5} \rightarrow 4_{0,4}$ & T$_{\rm MB}$ & - & - & 0.34$\pm$0.02 
       & $\leq$0.1 & 0.23$\pm$0.04 & $\leq$0.07 & 0.11$\pm$0.02\\
       & $\Delta v$ & - & - & 5.1$\pm$0.4 & - & 2.6$\pm$0.4 & - & 
       12.3$\pm$1.0\\
       & $\int T_{\rm MB} \delta v$ & - & - & 2.2$\pm$0.4 & $\leq$0.2 & 
       0.7$\pm$0.2 & $\leq$0.08 & 1.3$\pm$0.3 \\
SO$_{2}$ $5_{2,4} \rightarrow 4_{1,3}$ & T$_{\rm MB}$ & - & - & 0.23$\pm$0.02 
       & $\leq$0.09 & $\leq$0.09 & $\leq$0.07 & $\leq$0.07 \\
       & $\Delta v$ & - & - & 6.2$\pm$0.3 & - & - & - & - \\
       & $\int T_{\rm MB} \delta v$ & - & - & 1.2$\pm$0.2 & 
       $\leq$0.1 & $\leq$0.2 & $\leq$0.07 & $\leq$0.3 \\
\hline
\end{tabular}
\end{table*}

Figure~\ref{4maps} displays the integrated intensity maps of the SO
$3_{2} \rightarrow 2_{1}$ and $4_{3} \rightarrow 3_{2}$ lines, as well as the 
SO$_{2}$ $3_{1,3} \rightarrow 2_{0,2}$ and $5_{1,5} \rightarrow 
4_{0,4}$ lines, and, for comparison, the CS $2 \rightarrow 1$ and 
$3 \rightarrow 2$ lines \citep{1996ApJ...471L.111L}. 
Three peaks of emission stand out on the maps: one associated with the East lobe
of the outflow (marked E in the map - offset 72$''$, -12$''$) 
and two associated with the West lobe (marked W1 -offsets -60$''$, 12$''$-
and W2 -offset -96$''$, 24$''$- respectively).
On the contrary, no enhanced emission is observed towards the source itself 
with the exception of the CS $3 \rightarrow 2$ line. 
The E and W2 regions are symmetrically situated around IRAS2, and have 
previously been identified by \citet{1996ApJ...471L.111L}
\footnote{The CS $5\rightarrow 4$ line has also been observed but is 
seldon detected. Thus, we did not use it for our analysis.} as the bow 
shocks at the end of the East and West lobes of the outflow. The three  
positions E, W1 and W2 coincide with peaks of CH$_3$OH emission 
\citep{1998A&A...335..266B}. 
 
For a proper 
comparison of the spectra taken at different frequencies, all spectra
were smoothed to the same spatial resolution of $24''$, which corresponds to
the largest beam \citep[including the CS observations from]
[]{1996ApJ...471L.111L}. 
Note that the SO$_2$ $5_{1,5} \rightarrow 4_{0,4}$ and 
$5_{2,4} \rightarrow 4_{1,3}$ spectra at the IRAS2 position were not 
smoothed because of the lack of observed positions around the source.
However, since the target of the present study is the outflow and
not the source itself, this is not relevant.  
The spectra of the SO and SO$_{2}$ transitions observed towards 
IRAS2, E, W1 and W2 are shown in Fig.~\ref{spect},
and the derived line parameters are summarized in Table~\ref{intensity}. 

As already noticed, the emission of the SO and SO$_{2}$ 
lines is strongly enhanced in the E, W1 and W2 positions, whereas it is 
very weak 
toward the protostar itself. In the direction of IRAS2, the spectra 
exhibit profiles similar to the ones observed in the molecular 
cloud: very narrow lines with an intensity 
decreasing with the energy of the line suggesting that the emission is 
dominated by the cloud rather than the source.
Conversely, the spectra of the outflow regions are very intense
and exhibit high-velocity wings. 
In the western positions (W1 and W2 in Fig.~\ref{spect}), two velocity 
components can be clearly seen. 
The first component, with a broad profile, is blue-shifted by 6~km~s$^{-1}$ 
with respect to the velocity of the ambient cloud. 
We will refer to this component as the High Velocity Component, ``HVC''.
The second component peaks at the systemic velocity of the cloud, 7.6~km~s$^{-1}$. 
We will refer to this component as the Low Velocity Component, ``LVC''. 
While the HVC component clearly probes the outflowing gas, the situation is 
less clear 
for the LVC component because the maps are not large enough. 
However, there is some indication that the LVC component is associated with
the molecular cloud,  for the LVC component is extended and 
 is poorly contrasted (see Fig.~\ref{solowhigh}). Also, the v$_{LSR}$, 
7.6~km~s$^{-1}$, is constant across the emitting region. The lines of the 
East lobe (E) are brighter than those of the West lobe, and only one component 
is detected, red-shifted with respect to the systemic velocity. 
Most probably the component centered on the cloud velocity is buried 
of this red-shifted bright component, making it difficult to separate 
the two components. Similar profiles have been observed in the CS 
survey of NGC1333 by \citet{1996ApJ...471L.111L}.

Table~\ref{intensity} reports the velocity-integrated intensity in each 
position (the area under the Gaussian profile and integrated over the 
velocity interval at the ``zero-intensity'' level) which we will use 
to derive the gas temperature, density and species column densities. 
Note that we fitted separately the low (LVC) and high (HVC) velocity 
components. While this is easy for the West lobe emission where the two 
components are well separated, the fit is more tricky in the East lobe, where 
the two components are blended. In order to compute the area of the two 
components in the East lobe, we procede as follows. The LVC component is 
supposed to be centered on the systemic velocity of the cloud 
(7.6~km~s$^{-1}$) and to have a fixed width of 1~km~s$^{-1}$ (same width as 
the LVC component of the West lobe). The residual spectrum, after removing 
the LCV component, is HVC. Using this method, the resulting LVC integrated 
intensity is less than 30\% of the HVC one. Therefore, the uncertainty on HVC 
intensity produced by the mixing of the two components is low whereas the LVC 
one is less robust. For this reason, we will consider in the following only 
the HVC component of the East lobe.

Some but not all of the spectra of the West lobe can be fitted by two 
Gaussians,
representing the LVC and HVC components respectively.
On the contrary, none of the spectra of the East lobe can be analysed
by two Gaussians.
In those cases where the two Gaussian fit the line,
it would be more correct to consider the line intensity
of the LVC and HVC components computed
in this way, rather than just  integrating over  two different velocity
intervals, as explained above.
However, since the two Gaussian analysis is not possible everywhere,
we decided to adopt the same method for all the points.
While this choice does not practically affect the integrated intensity 
of the
HVC component (the errors remain within the observational errors),
it affects the LVC line intensity determination.
To give the order of magnitude of the possible error associated with the
adopted procedure, the two Gaussians analysis would give
at most a factor of 2  lower integrated intensity for the SO lines,
20\% less for the CS lines and no LVC component would be
detected in the SO$_2$ lines. The consequencies for the
computed LVC column densities and abundance ratios
(according to the analysis described in the next section and reported in
Tables~\ref{resuldiag} and \ref{colchi}) are the following: $N_{SO}$ would
be 30\% to 40\% lower, $N_{CS}$ would be 20\% lower and the SO$_2$/SO
abundance ratios would have an upper limit around 0.08 in the West LVC
whereas the CS/SO abundance ratios would be almost the same as in
Table~\ref{colchi}.
For this reason, our analysis will especially focus on the HVC emission,
where the determination of the integrated emission is robust.
On the contary, we will treat the LVC emission with more caution
in the rest of the paper.

%{\bf We derived the physical conditions (temperatures and H$_2$ densities), the SO, SO$_2$ and CS  column densities and the abundance ratios ($SO_{2}$/SO and CS/SO) at several positions of the two lobes of the outflow using both the rotational diagram method and a non-LTE LVG model. All these results are in Appendix~\ref{coldens_determination}.}

\section{Column densities and abundance ratios}\label{coldens_determination}
\begin{table*}
\caption{Column densities of SO, SO$_{2}$ and CS, as derived from an LVG
analysis, at several points along the outflow and on the source IRAS2 
position
The column densities (averaged on the 24$''$ beam) have been derived with a 
non-LTE LVG analysis (see text). As we only have one SO$_2$ transition 
detected on the source position, we used the same temperature and density as 
for CS to derive the SO$_2$ column density on IRAS2. LVC and HVC refer to the 
low and high velocity component respectively (see Sect.~\ref{results} for 
details). The uncertainties in the derived column densities are given 
by the observational errors, and are about 30\%. Consequently, the uncertainty 
in the abundance ratios is about 60\%. If we compute the LVC SO column 
densities with the LVG model using a temperature of 20~K and a density of 
$5\times 10^4$~cm$^{-3}$ instead of the $\chi^2$ method explained in 
section~5.2, we find column densities two times higher than the ones reported 
in this table.
\label{colchi}
}
\begin{tabular}{lcccccc}
\hline
\hline
$\Delta\alpha$, $\Delta\delta$ ('')  & N$_{\rm SO}$ & N$_{\rm SO_{2}}$ & 
N$_{\rm CS}$  & $SO_{2}$/SO & CS/SO \\
 & ($10^{14}$) & ($10^{13}$) & ($10^{13}$) & & \\
\hline
0 0 (IRAS2)  & 0.8 & 2.1 & 4.0 &   0.2 & 0.5 \\
48 -12 (HVC) & 1.5 & 2.5 & 0.4 &  0.2 & 0.03 \\
60 -12 (HVC) & 3.0 & 7.0 & 1.2 &  0.2 & 0.04 \\
72 -12 (HVC) & 3.5 & 7.0 & 2.6 &  0.2 & 0.07 \\
84 -12 (HVC) & 2.0 & 2.0 & 1.8 &  0.1 & 0.09  \\
-36 12 (LVC) & 0.9 & $\leq 0.5$ & 4.5 & $\leq 0.05$ & 0.5 \\
-36 12 (HVC) & 1.0 & 2.0 & 3.0 &  0.2 & 0.3 \\
-48 12 (LVC) & 0.6 & 1.2 & 3.8 &  0.2 & 0.6 \\
-48 12 (HVC) & 1.0 & 2.0 & 2.0 & 0.2 & 0.2 \\
-60 12 (LVC) & 0.9 & 1.2 & 3.7 & 0.1 & 0.4\\
-60 12 (HVC) & 1.5 & 2.5 & 2.5 & 0.2 & 0.2\\
-84 24 (LVC) & 0.6 & 0.8 & 2.5 &  0.1 & 0.4 \\
-84 24 (HVC) & 1.5 & 2.5 & 4.3 &  0.2 & 0.3 \\
-96 24 (LVC) & 0.7 & 0.7 & 2.7 &  0.1 & 0.3 \\
-96 24 (HVC) & 2.0 & 6.5 & 2.9 &  0.3 & 0.1 \\
\hline
\end{tabular}
\end{table*}
\begin{figure}
\centering
\includegraphics[angle=90,width=0.8\linewidth]{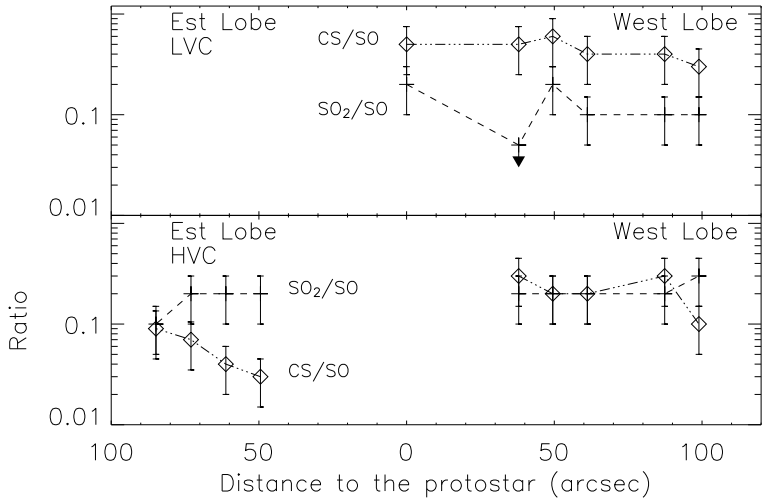}
\caption{SO$_2$/SO and CS/SO abundance ratios (Table~\ref{colchi}) as a 
function of the distance to the protostar. Upper panel represents the LVC 
ratios of the West lobe (SO$_2$/SO ratios are upper limits). Lower panel 
shows the ratios in the HVC of East and West lobes. Note that the uncertainty 
in the values of the ratios is about 60\%.}
\label{cs_so_so2_d}
\end{figure}

We estimated the column densities of the SO, SO$_2$ and CS molecules and put some constraints on the physical conditions (temperature and H$_2$ density) at selected positions along the outflow. Both the rotational diagram method and a non-LTE LVG code were used. In this section, we present only a summary of the results. The details of this analysis are given in Appendix~\ref{append_coldens}. 

For the LVC component, our modeling suggests densities between $2\times 10^4$ and $2\times 10^5$~cm$^{-3}$ and temperatures about 25~K. Thus for the modeling of the chemistry (next section), we will assume a density of $5\times 10^4$~cm$^{-3}$ and a temperature of 20~K for this component. For the HVC component, we found a difference in the physical conditions between the two lobes of the outflow: the temperatures and densities derived from SO$_2$ in the East lobe suggest a gradient along the outflow. We also found a slight difference between SO and SO$_2$. Indeed, SO seems to probe a gas less dense that the gas probed by the SO$_2$ transitions, suggesting a spatial differentiation in the formation of these two molecules. However, the data are not sufficient to better quantify this, so that in the following we will assume in our chemical modeling that both molecules originate in the same gas, where the density is around $10^5$~cm$^{-3}$ and the temperature is about 50~K, the median values between SO and SO$_2$ derivations.

The column densities derived using both methods differ by less than 30\%. The column densities and the SO$_2$/SO and CS/SO abundance ratios obtained with the LVG model are summarized in Table~\ref{colchi} for the several positions in the outflow. The abundance ratios are shown as a function of the distance to the protostar in Fig.~\ref{cs_so_so2_d}. Remarkably, the derived abundance ratios are different in the LVC and HVC
components, which, therefore, differ not only in the kinematics and
gas temperature and density, but also in the chemical composition.

In the next section we will try to understand the differences in the chemical
composition of the LVC and HVC components, as well as the gradients
observed along the outflow, comparing
the observed ratios with the theoretical ratios of the 
\citet{2004A&A...422..159W}  chemical model, predicted for the physical 
conditions described above.

\section{Chemical model}\label{chem_mod}

\begin{table}
\centering
%\begin{small}
\caption{Initial conditions for the chemical modelling. $^a$ Elemental abundances used to compute the gas phase composition in molecular clouds (step 1). $^b$ Species contained in grain mantles and sputtered in the gas phase in shocks (step 2). We took the abundances observed on grains in high mass protostars regions. $^c$ The cosmic abundance of sulphur is $3\times 10^5$ \citep{1994ApJ...430..650S}. We used a 30 times lower abundance to take into account the high depletion of sulphur in molecular clouds \citep{1994A&A...289..579T,2004A&A...422..159W}. $^d$ We varied the amount of sulphur sputtered from the grain mantle (see text).
Ref: (1) \citet{1994ARA&A..32..191W}; (2) \citet{1996ApJ...467..334C}; (3) \citet{1998ApJ...493..222M}; (4) \citet{1996A&A...315L.333S}; (5) \citet{2001A&A...376..254K}; (6) \citet{1996ApJ...472..665C}; (7) \citet{1998A&A...336..352B}; (8) \citet{1997ApJ...479..839P}; (9) upper limit from \citet{1998ARA&A..36..317V}. \label{CI}}
\begin{tabular}{lcc}
\hline
\hline
\multicolumn{3}{c}{Elemental abundances$^a$} \\
\hline
Species & Abundance & Ref. \\
He & 0.28 & (1) \\
O & $6.5 \times 10^{-4}$ & (2) \\
C$^{+}$ & $2.8 \times 10^{-4}$ & (3) \\
S$^{+}$$^c$ & $1.0 \times 10^{-6}$ &  \\
\hline
\hline
\multicolumn{3}{c}{Grain mantle composition$^b$}\\
\hline
Species & Abundance & Ref.  \\
\hline 
H$_{2}$O &$10^{-4}$ & (4)  \\ 
H$_{2}$CO & $4 \times 10^{-6}$ & (5)  \\ 
CH$_{3}$OH & $4 \times 10^{-6}$  & (6)  \\
CH$_{4}$ & $10^{-6}$  & (7) \\ 
OCS & $10 ^{-7}$ & (8) \\
H$_2$S & $10 ^{-7}$ & (9) \\
S$^d$ & $3\times 10^{-7} - 3\times 10^{-5}$ & \\
\hline
\end{tabular}
%\end{small}
\end{table}

In this section, we compare the abundance ratios found in the LVC and HVC
components with the predictions the chemical model we
developed recently to study the sulphur chemistry \citep{2004A&A...422..159W}.
Briefly, the model  computes the evolution of the chemical composition
of a gas
at a fixed temperature and
density, given an initial composition of the gas phase and of the grain
mantles
components released in the gas phase at time=0.
In this sense, it is a pseudo time-dependent model, for it does not take
into
account the physical evolution of the gas, but only the chemical evolution.
For the specific case of the molecular shocks we model here,
the shock is assumed to warm and compress the gas, giving rise to a sudden
increase of the gas temperature and density. Also, the grain
mantles are sputtered by the ions accelerated in the shock (e.g. Pineau
de Forets et al.
1993) and/or destroyed by the collisions between grains \citep[e.g.][]{1994ApJ...433..797J}.
As a result, some or all the grain mantle components are injected into
the gas
phase at the shock passage. The model thus computes the gas chemical
composition
as a function of time, following those changes: gas density and
temperature, as well as
the gas chemical composition due to the grain mantle (partial) evaporation.

Note that the model does not aim to reproduce all the
molecular abundances, but it specifically focuses on the sulphur chemistry.
In this respect, it has the most up-to date and complete set of reactions
involving S-bearing molecules.
%, and can be considered the status of the art on this specific issue.

In practice, in the present study the passage of the shock is simulated
by a two-step
process:\\
1) Before the shock, the gas composition is computed for a cloud with
the density and
temperature derived by the observations of the LVC, namely $5\times 10^{4}$~cm$^{-3}$ and 20 K. The elemental abundances used for this step are summarized in Table~\ref{CI}. 
The age of the cloud is estimated by comparing the model predictions
with the observed CS/SO and SO$_{2}$/SO abundance ratios.\\
2) The cloud chemical composition computed in step 1 is used as the initial
condition for step 2 computations.
To the cloud composition, we add the grain mantle components released
into the gas phase (see Table~\ref{CI}), and then the model follows the chemical evolution of 
the gas at $0.1-10\times 10^{5}$ cm$^{-3}$ and 25-120 K
(namely a range of temperatures and densities appropriate for the 
conditions found in the HVC). Following the results of \citet{2004A&A...422..159W}, we assume that sulphur mainly evaporates in the H$_2$S, OCS and atomic form and we vary the amount of atomic sulphur released in the gas phase. Note that we also considered the possibility that sulphur evaporates in the molecular form S$_2$ but as in the case of the hot cores \citep{2004A&A...422..159W} the model with the evaporation of S$_2$ does not reproduce the observations in shocks. We will therefore not discuss this case further. \\

In summary, we studied the influence of four parameters: the temperature, the density, the time since grain mantle sputtering and the amount of atomic sulphur released from the grains.

\subsection{Results: the cloud gas (LVC component)}\label{mod_cold}
\begin{figure}
\centering
\includegraphics[width=0.7\linewidth]{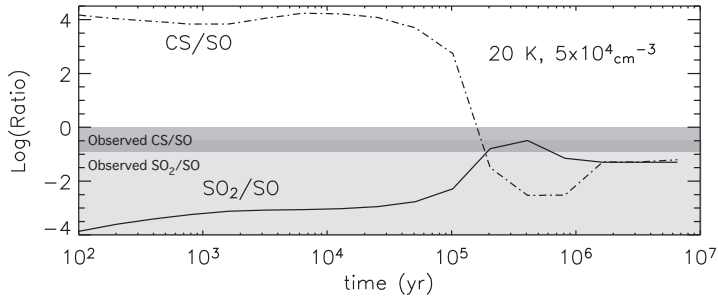}
\caption{Theoretical SO$_2$/SO and CS/SO abundance ratios as a function 
of time.
The temperature is 20~K, and the density  $5\times 10^4$~cm$^{-3}$.
The two grey bands represent the observed ratios in the LVC component of
the East lobe
(maximum and minimum values from Table~\ref{colchi} with the uncertainty of 50\%).
Note that for the SO$_2$/SO ratio, we only have a upper limit in the -36$''$, 12$''$ LVC position.
The darker band corresponds to the overlap of the observed values. }
\label{Cloudiras2}
\end{figure}
  Fig.~\ref{Cloudiras2} shows the SO$_2$/SO and CS/SO abundance ratios as predicted by the chemical model for the LVC temperature and density (20~K and $5\times 10^4$~cm$^{-3}$) as a
function of time. The observed ratios are reported on the same figure. The comparison between the model predictions and the observed ratios indicate that the cloud age is around $2\times 10^5$~yr or greater than about $10^6$~yr.
 In that case, the observed SO$_2$/SO ratios are in agreement 
 with the model whereas the predicted CS/SO ratios are four times lower 
 than the observed ones. An increase of the density or a decrease of the 
 temperature leads to a better agreement between observed and predicted ratios. 
 A decrease of the density or an increase of the temperature would have the 
 opposite effect. For an age of $2\times 10^5$~yr, the model predicts absolute 
 abundances 10 times higher than the observed ones whereas the observed 
 abundances are reproduced by the model within a factor of 3 for times longer 
 than $10^6$~yr. This analysis assumes that SO, SO$_2$ and CS all originate in
 the same gas, which in the case of the molecular cloud is likely a good
 approximation. As shown in Fig.~\ref{Cloudiras2}, the CS/SO and SO$_2$/SO
 ratios are each clocks that work differently depending on the temperature and
 density. Indeed, because of the different dependence on the different model
 parameters, the use of both ratios can help constrain the age along with the
 density and temperature of the gas, as discussed above.

\subsection{Results: the shocked gas}
\begin{figure}
\centering
\includegraphics[width=0.7\linewidth]{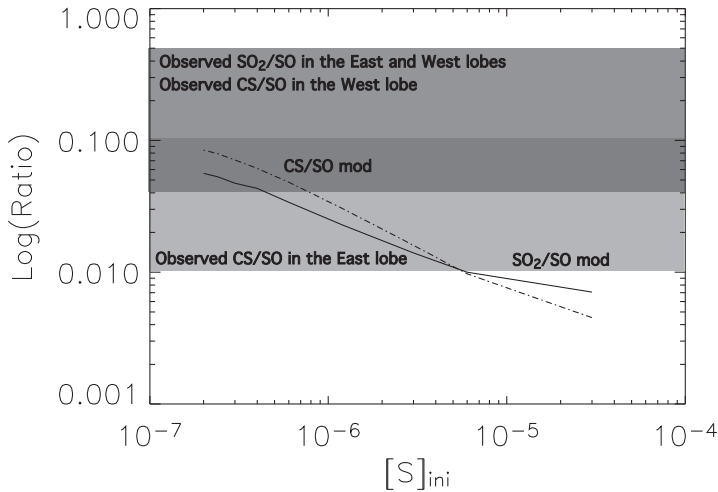}
\caption{Theoretical SO$_2$/SO (solid line) and CS/SO (dash dot line) abundance ratios as a function
of the initial amount of atomic sulphur evaporated in the gas phase. We varied the amount of atomic sulphur (abundance compared to H$_2$) evaporated in the gas phase between $3\times 10^{-7}$ and  $3\times 10^{-5}$. All the other model parameters are fixed: T=50~K, n(H$_2$)=$10^5$~cm$^{-3}$ t=2000~yr. The two grey bands represent the observed ratios in the shocked gas of 
the East and West lobes (see Table~\ref{colchi}). Note that the two bands are superimposed in the middle (Log(Ratio)=0.004-0.1). }
\label{iras2Si}
\end{figure}
\begin{figure}
\centering
\includegraphics[width=0.8\linewidth]{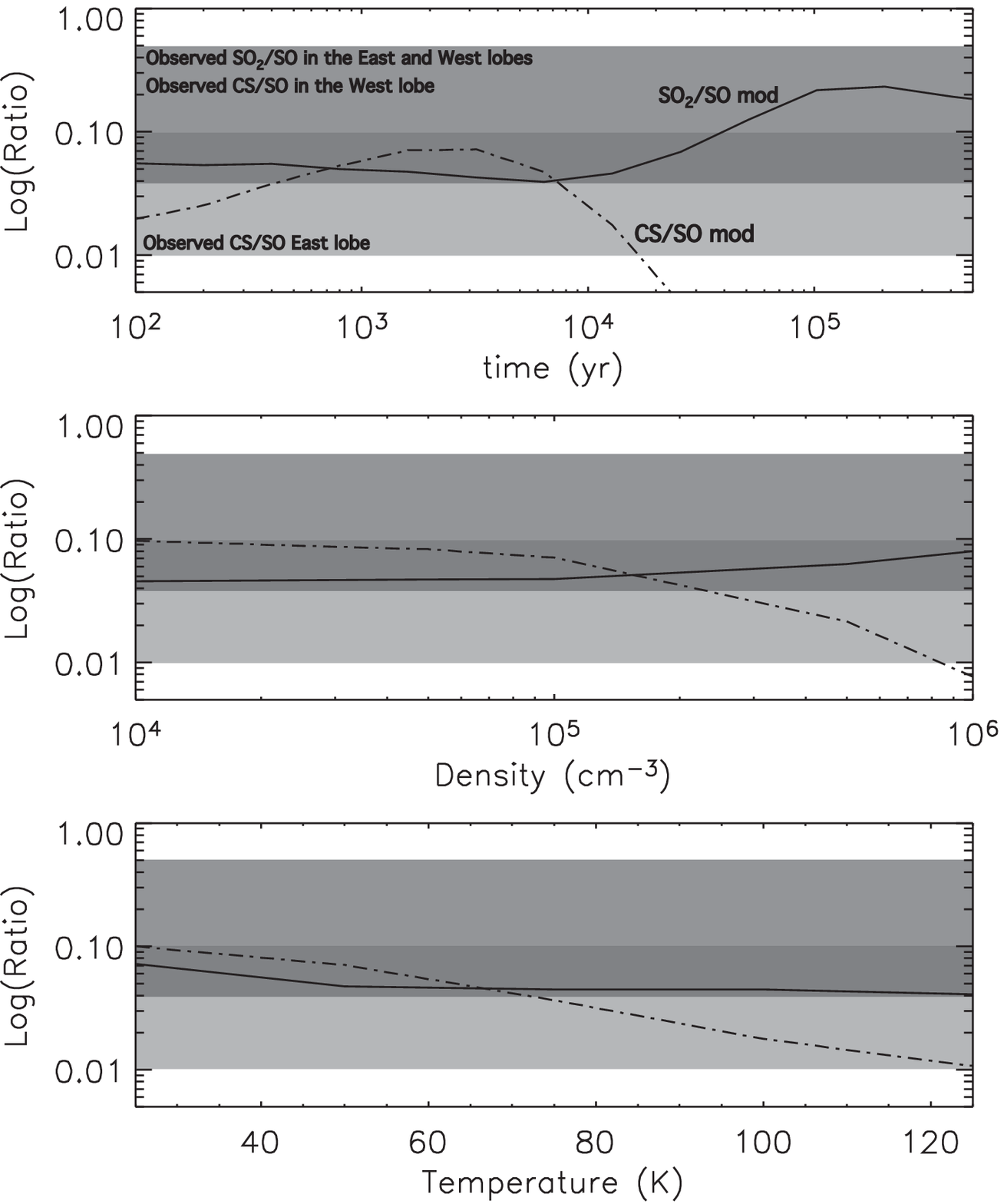}
\caption{Theoretical SO$_2$/SO and CS/SO abundance ratios as a function
of three parameters: time (upper panel), the H$_2$ density (middle panel) and the
temperature (lower panel). In each panel, the other parameters are fixed: T=50~K and n(H$_2$)=$10^5$~cm$^{-3}$ in the upper panel, T=50~K and t=2000~yr in the middle panel, and 
n(H$_2$)=$10^5$~cm$^{-3}$ and t=2000~yr in the lower panel. For all the figures, the initial amount of
evaporated atomic sulphur is  $3\times 10^{-7}$. The two grey bands represent the observed ratios
in the shocked gas of the East and West lobes.(see Table~\ref{colchi}). Note that the two bands are superimposed in the middle (Log(Ratio)=0.004-0.1). }
\label{Modiras2}
\end{figure}

As we already mentioned, in addition to the temperature, the density and the time, we studied a fourth parameter: the amount of atomic sulphur injected in the gas phase which depends on the shock strengh. This last parameter, the amount of atomic sulphur, is expected to be the same across the outflow. Consequently, we start studying the dependence of the predicted abundance ratios on this parameter. We took the average density and temperature previously derived by the SO emission in the HVC, namely $10^5$~cm$^{-3}$ and 50~K respectively. Anticipating the results on the time dependence of the predicted ratios, we took an early time, 2000~yr. For these conditions, we ran several models, varying the initial amount of atomic sulphur. 
Fig.~\ref{iras2Si} shows the results of the modelling together with
the range of observed ratios in the two lobes of the outflow. The
first robust result is that only a small fraction of the sulphur is
evaporated in the gas phase: $\le 1/100$ of the elemental sulphur in
the West lobe and $\le 1/40$ of the elemental sulphur in the East
lobe. This suggests that the shocks in IRAS2 are rather mild. The
second result is that H$_2$S and OCS evaporate from the grain more easily than 
atomic sulphur. Indeed, decreasing the OCS injected in the gas by the same amount as S would give predictions not in agreement with the observations, because the abundance of CS in the warm shocked gas highly depends on the initial abundance of OCS. The decrease of H$_2$S does not influence the results significantly. We will discuss this point further in Sect.~7.

Taking the above composition for the injected material, we explored the
dependence of the SO$_2$/SO and CS/SO abundance ratios as function of the
other three parameters of the model: the density, temperature and age of
the shock. The results are shown in Fig.~\ref{Modiras2}. 

The SO$_2$/SO ratio depends weakly on the shock age, until $10^4$~yr, and on the density and temperature of the shocked gas. On the contrary, the CS/SO ratio increases with time for $(2-3)\times 10^3$~yr before decreasing sharply because SO is efficiently produced whereas the abundance of CS does not change. The CS/SO also decreases with the density, for $n_{H_2} \ge 10^5$~cm$^{-3}$ and with the temperature. The CS/SO ratio is therefore an efficient constraint on the shock age, temperature and density. We will discuss the constraints derived along the outflow in Sect.~7.

Finally, we notice that the HVC chemical modelling depends on the initial gas 
phase composition computed in step 1 (see Sect.~\ref{chem_mod}), because the 
gas temperature (50~K) is too low to reset the chemistry as is the case for the 
hot cores. In addition, the CS/SO ratio depends more on the initial composition 
than the SO$_2$/SO ratio because CS is also directly influenced by the initial 
abundance of atomic carbon (wich is a function of the cloud age). Indeed,
the initial abundances are parameters hard to constrain, which we decided not
to alter. The absolute 
abundances depend even more on the model parameters than the abundance ratios. 
Indeed, we are not able to reproduce the observed absolute abundances which 
are two orders of magnitude higher than the modelled ones. This discrepancy can
have various causes. The first is that the modelling is wrong, and only further observations will be able to rule out this possibility. Other
possibilities include the uncertainty on the adopted density and temperature.
Here we are using average values across each modeled point, but there may be a stratification of density, temperature and chemical
composition, that may affect the results. Also, the derived absolute abundances are rather uncertain because they
are based on observations of low J CO transitions and this may be
the most important cause of the discrepancy between the model and the
observations. These
observations have relatively large beams, and the amount of gas probed by the CO
lines can be substantially larger than the gas where the SO, SO$_2$ and CS lines
are emitted, even up to a factor of 100. Only high spatial resolution observations will be able
to clartify this possibility. In addition, the CO 2-1 line used to
derive the H$_2$ column density may be optically thick, underestimating the
H$_2$ column density, and only $^{13}$CO observations will be able to clarify
this point. In the absence of such observations we think that our modelling, based
on the molecular abundance ratios -which are more robust than the absolute
abundances- gives a reasonable and consistent interpretation of the
observations. We will try to explore in detail the consequences of this
interpretation in the next section.

%The problem can be 
%the averaged physical conditions we are using for the chemical modelling which 
%influence mostly the absolute abundances. However, the goal  of this work is 
%to understand the difference between the East and West lobe by roughly 
%reproducing the sulphur bearing behaviour and not to make a rigorous chemical 
%modelling of the shocks. 

\section{Discussion}

\subsection{Nature of the LVC component}

The density and temperature derived in the LVC (Sect.~\ref{Mod_resul}) are consistent with the hypothesis suggested in \S 4,
that the LVC is associated with the molecular cloud. Indeed, observations of the extended $^{12}$CO and $^{13}$CO emission suggest a density of $\sim 10^5$ cm$^{-3}$ and temperature of $\sim 20$ K for the gas cloud \citep[see][]{1996A&A...306..935W,2003ApJ...582..830B}. Likely, this gas is rather in  the warm layers just behind the PhotoDissociation Region created by the illumination
from a G$_0 \sim 100-400$ FUV field \citep[see discussion in][]{2003ApJ...582..830B}. The SO column density in the LVC component is about $5-9 \times 10^{13}$~cm$^{-2}$, which would translate into a SO abundance of about
$0.6-1 \times 10^{-9}$, assuming that the warm layers extend for about $8\times 10^{22}$ cm$^{-2}$ as derived by  \citet{1996A&A...306..935W}.
This SO abundance is consistent with determinations in other molecular
clouds \citep{1989ApJ...345..828S,1995PASJ...47..845H} and the modeling of them \citep{2004A&A...422..159W}, so that we are tempted to conclude that the LVC component is dominated by the cloud emission.

Also from a chemical point of view, the SO$_2$/SO and CS/SO ratios
observed in LVC seem to reflect the cloud chemistry, even though the
chemical model predicts a CS/SO ratio lower than what is observed. 
%This can be explained by a density higher than $5\times 10^4$~cm$^{-3}$. Indeed, using other molecules \citet{2004A&A...415.1021J} derived the same temperature but a higher a density of $10^6$~cm$^{-3}$. The density we derived in LVC is consistent with the density of HVC which can not be as high as $10^6$~cm$^{-3}$ (see next section), i.e. that the density in LVC is lower than in HVC. 

Finally, the constraints on the cloud age given by the sulphur chemistry, i.e. $\sim 2\times 10^5$~yr or $\ge 10^6$~yr, are remarkably in agreement with previous estimates from \citet{1996AJ....111.1964L}, $\le(1-2)\times 10^6$~yr, obtained by analysing Near-IR survey data. Therefore, the cloud hosting IRAS2/NGC1333 is chemically young, less than $10^6$~yr old.

\subsection{Age and physical conditions of the outflow}

The comparison between the observed abundance ratios and the chemical modelling presented in Sect.~6 gives some constraints on the shock age, density and temperature. First, the sulphur bearing species indicate an age of $5\times 10^2 - 7\times 10^3$~yr for the outflow of IRAS2. This age is in agreement with the range estimated by \citet{1998A&A...335..266B}, $4\times 10^2 - 5\times 10^3$~yr, based on the dynamics of the bullets of the IRAS2 outflow. The estimated age of the outflow is also in agreement with the general scheme of formation and destruction of molecules in shocks proposed by \citet[][, Table 7]{2004A&A...422..159W}: the simultaneous presence of SO and SO$_2$ would indicate a relatively young shock ($\le 10^3$~yr). The accuracy on the SO$_2$/SO and CS/SO ratios is however not good enough to constrain the age of each clump and therefore to test if they are different events.
The density of the HVC component in the West lobe is constrained by the CS/SO ratios to be lower than $2\times 10^5$~cm$^{-3}$ and the temperature lower than 70~K. These physical conditions are in remarkable agreement with the non-LTE analysis of the SO emission presented in Sect.~5 (Table~\ref{phys_chi}). No constraints on the temperature or density can be deduced for the HVC component in the East lobe. However, the chemical modelling shows that an increase of the density and/or of the temperature leads to smaller CS/SO ratios. Thus, the lower CS/SO ratios observed in the East lobe compared with the West part of the outflow can be explained by a denser and/or warmer gas. The study of molecular excitation conditions (Table~\ref{phys_chi}) seems to confirm that the gas in the East lobe is denser than in the West lobe. 

The density and temperature we derived in HVC are lower than that obtained
by \citet[][70~K and $10^6$~cm$^{-3}$]{2004A&A...415.1021J}. For the
temperature, we are still within the mutual uncertainties. For the density 
value, we are forced to conclude that $10^6$~cm$^{-3}$ is
not in agreement with the HVC SO and SO$_2$ emission data since a high
density would require gas temperatures too low, similar to the
rotational temperatures (in Table~\ref{resuldiag}). Indeed, we showed
that optically thick lines cannot explain these low rotational
temperatures (see Annex). One possible explanation for the different
results by \citet{2004A&A...415.1021J} is that they did not consider the two lobes of the outflow separately, and also they treated all the
studied molecules without studying the excitation conditions of each
molecule individually. As we showed, the physical conditions are different in the
East and West lobes, and different molecules probably trace different
layers of shocked gas \citep[see also][]{2001A&A...372..899B}. In any case, the SO and CS densities computed by \citet{2004A&A...415.1021J} are similar to ours in both LVC and HVC.

The low temperature ($\le 70$~K) derived in the HVC component from SO and SO$_2$ emission, and from other molecular emission by \citet{2004A&A...415.1021J}, as well as the small fraction of sulphur released in the shocked gas found from the chemical modelling suggest mild shocks. The SO, SO$_2$ and CS molecules seem rather to trace the entrained material, mildly shocked, rather than the shock from the outflowing wind. This would explain both the lack of gas compression and the low gas temperature.

\subsection{Consequences for the depleted form of sulphur}

In this work on sulphur chemistry in shocks, we assumed that sulphur was sputtered from the grains in the OCS, H$_2$S and S forms. This assumption does not imply that sulphur is depleted in the atomic form but that the third form of depleted sulphur is quickly converted into S once evaporated in the gas phase. 
We denote this species {\bf S?}.In \citet{2004A&A...422..159W}, we proposed that S? is a polymeric form of sulphur such as S$_8$. 

One conclusion of this work is that it is more difficult to sputter S? than OCS and H$_2$S. The efficiency of each molecule sputtering depends on the binding energy of that molecule. Thus S? should have a higher binding energy than OCS and H$_2$S. Unfortunatly, this parameter is poorly known and depends highly on the grain surface composition.

\section{Conclusions}

We studied the SO, SO$_2$ and CS emission in the protostar IRAS2 region and its East-West outflow. Using both LTE and non-LTE methods, we computed the species column densities of the gas associated with the outflow and the ambient medium respectively. The computed CS/SO and SO$_2$/SO abundance ratios were compared with the predictions of a chemical model. \\
The results are:
\begin{itemize}
\item[-] The sulphur bearing emission is similar in the IRAS2 position and in the ambient medium whereas it is enhanced in the shocked regions at the interface between the cloud and the wind.
\item[-] The low velocity component (LVC) of the sulphur bearing spectra reflects the physical conditions and the chemistry of the cloud younger than $10^6$~yr.
\item[-] The chemical modelling suggests that the IRAS2 outflow is
  young ($\le 2\times 10^3$~yr) but the accurary of the observed
  abundance ratios is not high enough to differentiate in time each clump of the outflow. The density of the gas traced by SO and SO$_2$ is lower than $2\times 10^5$~cm$^{-3}$ and the temperature lower than 70~K in the West lobe. The gas density in the East lobe is higher. 
\item[-] The theoretical modeling suggests than SO and SO$_2$ trace different layers of gas and the entrained material rather than the shock itself. Thus, these molecules are not appropriate to study the physical conditions of the gas and the evolution of the outflow. CS on the contrary seems to be more sensitive to these parameters and the use of CS/SO ratios can help in understanding the molecular shocks associated with the outflows.
\item[-] Sulphur-bearing species depleted on the grain
  mantles may be more difficult to sputter than OCS and H$_2$S. Thus, this species should have a higher binding energy than OCS and H$_2$S. 
\end{itemize}

\begin{acknowledgements}
We thank the IRAM staff in Pico Veleta for their
assistance with the observations, and the IRAM Program
Committee for their award of observing time. V.Wakelam wishes to thanks Guillaume Pineau des For\^ets for helpful discussions.

\end{acknowledgements}

%%%%%%%%%%%%%%%%%%%%%%%%%%%%%%%%%%%%%%%%%%%%%%%%%%%%%%%%%%%%%%%%%%%%%%%%%%%%%%
%%%%%%%%%%%%%% APPENDIX %%%%%%%%%%%%%%%%%%%%%%%%%%%%%%%%%%%%%%%%%%%%%%
\appendix

\section{Determination of the column and physical densities} \label{append_coldens}

In this section we report the details on the estimates of the column densities of
the SO, SO$_2$ and CS molecules at selected positions along the ouflow.
We used both the rotational diagram method and a non-LTE
LVG code. The two methods and the relevant results are discussed
separately in the following. We anticipate that they give 
values for the column densities that differ by less than 30\%.

\subsection{Rotational diagram method}\label{LTE}
\begin{figure}
\includegraphics[angle=90,width=15cm]{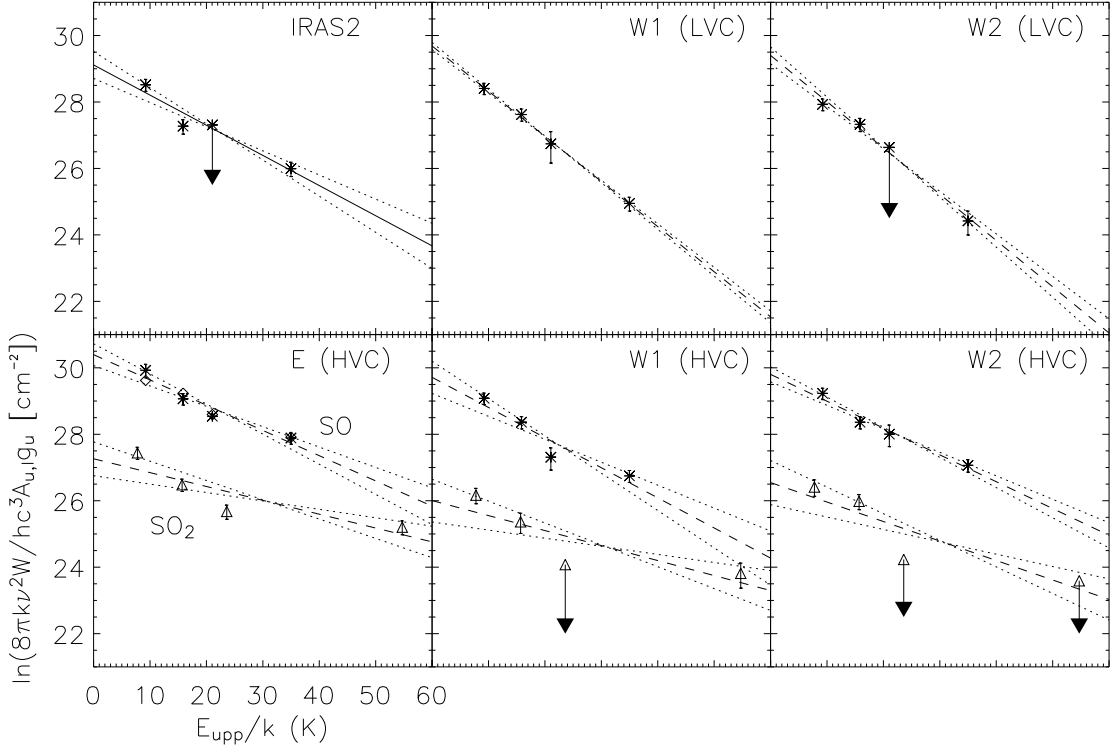}
\caption{Rotational diagrams of the SO and SO$_2$ molecules in the 
direction of IRAS2, E, W1 and W2 regions. In W1 and W2 we have constructed
separate diagrams for the LVC and HVC components: W1(LVC), W1(HVC), W2(LVC) 
and W2(HVC) whereas in E, we give only the HVC component 
(see text). 
Stars and triangles correspond to the SO and SO$_2$ observed 
data respectively. 
%The diamonds trace the values computed with an LVG model using the 
%column densities, temperature and densities derived from the chisquare 
%method and summarized in Table~\ref{colchi}(see text).
}
\label{diagrot}
\end{figure}

As widely discussed in the literature \citep[e.g.][]{1999ApJ...517..209G}, the 
rotational diagram method rests on the assumption that the molecular levels 
are in local thermal equilibrium (LTE) and that the lines are optically thin. 
Using the four observed transitions of SO and SO$_2$, we 
constructed rotational diagrams at the three positions
 E, W1 and W2, and at the IRAS2 position (see Fig.~\ref{diagrot}). 
Note that we did not perform this analysis for the CS data because we have
only two transitions available.
The obtained SO and SO$_2$ rotational temperatures 
and column densities are reported in Table~\ref{resuldiag}. 

\begin{table*}
\caption{Rotational temperatures and column densities of SO and SO$_2$ 
computed from the rotational diagrams of Fig.~\ref{diagrot}.
Note that the detected lines of SO$_2$ transitions towards IRAS2, 
W1(LVC) and W2(LVC) are not numerous enough to 
construct the SO$_2$ rotational diagram in these regions. Thus the 
SO$_2$ column densities towards IRAS2 and the LVC components (W1 and W2 LVC) 
are derived assuming LTE and an excitation temperature of 10~K.
\label{resuldiag}}
    \begin{tabular}{lccccc}
\hline
\hline
 & \multicolumn{2}{c}{SO} & \multicolumn{2}{c}{SO$_2$} & CO \\
$\Delta\alpha$, $\Delta\delta$ ($\arcsec$) & $T_{rot}$(K) & 
$N_{\rm SO}$($10^{14}\rm cm^{-2}$) & $T_{rot}$(K) & 
$N_{\rm SO_2}$($10^{13} \rm cm^{-2}$) & $N_{\rm CO}(10^{17}$cm$^{-2}$) \\
\hline
0, 0 (IRAS2) & $11 \pm 3$ & $0.9 \pm 0.01$ & 10 & $0.5 \pm 0.2$ & \_ \\
72, -12 (E HVC) & $13\pm 3$ & $4.0 \pm 0.4$ & $28 \pm 11$ & $10 \pm 1$ & 
0.75 \\
-60, 12 (W1 LVC) & $7.3 \pm 0.2$ & $0.8 \pm 0.04$ & 10 & $0.3 \pm 0.07$ & \_ \\
-60, 12 (W1 HVC) & 11.7$\pm$2.8 & 1.6$\pm$0.2 & 28.1$\pm$12.9 & 2.8$\pm$0.2 & 
0.45 \\
-96, 24 (W2 LVC) & $7.2 \pm 0.6$ & $0.6 \pm 0.07$ & 10 & $0.2 \pm 0.07$ & \_ \\
-96, 24 (W2 HVC) & 12.5$\pm$1.6 & 2.6$\pm$0.1 & 19.6$\pm$7.1 & 3.2$\pm$0.3 & 
\_ \\
\hline
\end{tabular}
\end{table*}

As previously found for CH$_3$OH 
\citep{1994A&A...285L...1S,1998A&A...335..266B}, the rotational temperatures 
are relatively low when compared to the expected temperatures in the shocked 
gas.
\citeauthor{1994A&A...285L...1S} concluded that IRAS2 was the first
example of a cold shock, while \citeauthor{1998A&A...335..266B} proposed
that the CH$_{3}$OH lines are subthermally excited. 
Indeed, the gas temperature in the IRAS2 outflow is certainly higher 
(i.e. $\geq 60~K$) than the rotational temperatures in Table \ref{resuldiag} 
since bright NH$_{3}$(3,3) emission has been detected towards the two
lobes \citep[see ][]{1998A&A...335..266B}. 
We therefore endorse the interpretation by \citeauthor{1998A&A...335..266B},
and we demonstrate that this is also the case for the SO and SO$_2$ molecules
in Appendix \ref{exc_cond}: the low rotational temperatures are simply
due to non-LTE effects.

\subsection{Non-LTE LVG method}\label{non-LTE}

\begin{figure}
\centering
\includegraphics[angle=90,width=16cm]{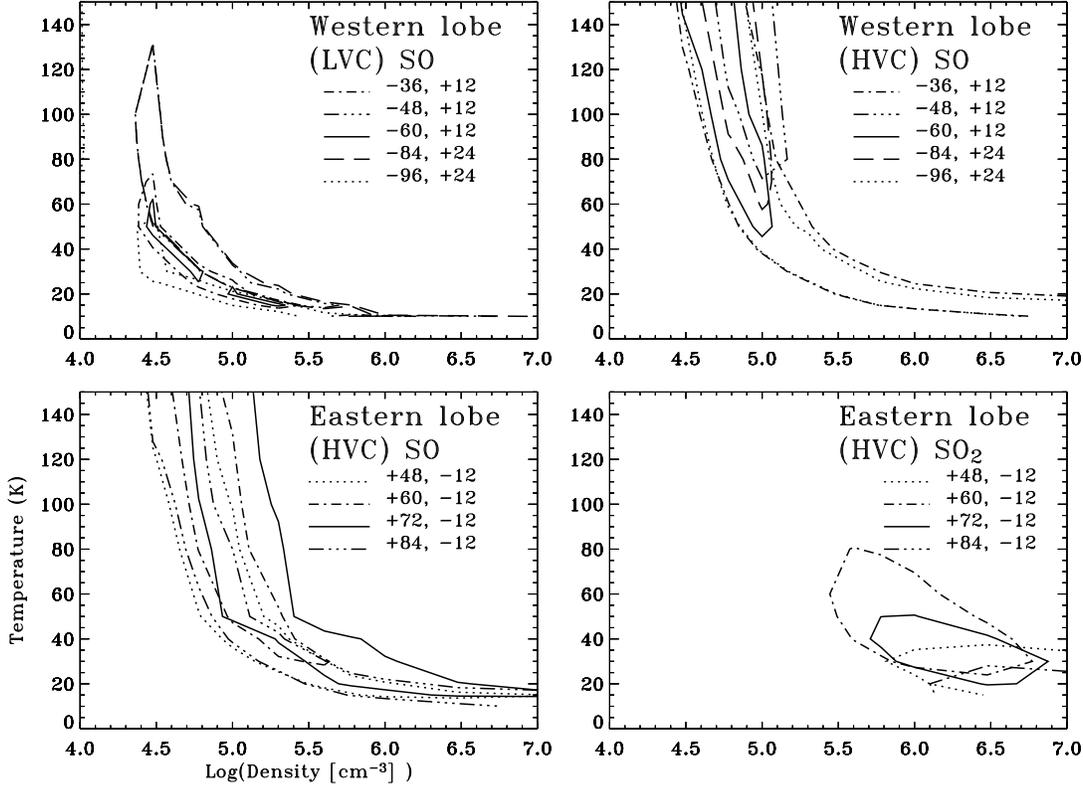}
\caption{$\chi^2$ contour at 2$\sigma$ (95.4\% in confidence) of the LVC and 
HVC SO line emission in the 5 positions of the West lobe (upper panels) and 
the HVC SO and SO$_{2}$ line emission in the 4 positions in the East lobe 
(lower panels) compared to the non-LTE LVG code.  }
\label{chisq}
\end{figure}
\begin{table*}
\caption{Temperatures and densities constrained by the SO and SO$_2$ 
observed line emission 
using a non-LTE LVG method (see text) for the different positions along 
the outflow. The last two columns summarise the temperatures and densities 
used to compute the CS column densities reported in Table~\ref{colchi}.  
\label{phys_chi}}
\begin{tabular}{l|cc|cc|cc}
\hline
\hline
&   \multicolumn{2}{c}{SO} &  \multicolumn{2}{|c}{SO$_2$}  &  \multicolumn{2}{|c}{Adopted for CS}  \\
$\Delta\alpha$, $\Delta\delta$ ($''$)   & T (K) & n (cm$^{-3}$) & T (K) & n 
(cm$^{-3}$) & T (K) & n (cm$^{-3}$) \\
\hline
0 0 (IRAS2) & - & $\geq 2\times 10^4$ & - & - & 20 & $5\times 10^4$ \\
48 -12 (HVC)    & - & $\ge 2.5\times 10^4$ & $\leq 35$ & 
$\geq 6.3\times 10^6$ & 50 & $10^5$\\
60 -12 (HVC)    & $\geq 30$ & $3.1\times 10^4$ - $4.0\times 10^5$ & 20-80 & 
$2.5\times 10^5$ - $6.3\times 10^6$ & 50 & $10^5$ \\
72 -12 (HVC)    & - & $\geq 5\times 10^4$ & 20-50 & 
$5\times 10^5$ - $7.9\times 10^6$ & 50 & $10^5$\\
84 -12 (HVC)   & - & $\geq 2.5\times 10^5$ & $\leq 30$ & $\geq 1.2\times 10^6$ & 50 & $10^5$\\
-36 12 (LVC)  & 14 - 75 & $2.2\times 10^4$ - $2.5\times 10^5$ & - & - & 20 & $5\times 10^4$\\
-36 12 (HVC) &  - & $\geq 2.5\times 10^4$ & - & - & 50 & $10^5$\\
-48 12 (LVC)  & 14 - 130 & $2.2\times 10^4$ - $5.0\times 10^5$ & - & - & 20 & $5\times 10^4$ \\
-48 12 (HVC) & $\geq 70$ & $4\times 10^4$ - $1.6\times 10^5$ & - & - & 50 & $10^5$ \\
-60 12 (LVC)  & 15 - 24 & $1.0\times 10^5$ - $2.2\times 10^5$ & - & - & 20 & $5\times 10^4$\\
& $\geq 25$ & $2.8\times 10^4$ - $6.3\times 10^4$ & & & & \\
-60 12 (HVC) & $\geq 45$ & $2.5\times 10^4$ - $1.2\times 10^5$ & - & - & 50 & $10^5$\\
-84 24 (LVC)  & $\le 130$ & $\geq 2.2\times 10^4$ & - & - & 20 & $5\times 10^4$\\
-84 24 (HVC) &  $\geq 60$ & $2.8\times 10^4$ - $1.0\times 10^5$ & - & - & 50 & $10^5$\\
-96 24 (LVC)  & $\le 55$ & $\geq 2.5\times 10^4$ & - & - & 20 & $5\times 10^4$ \\
-96 24 (HVC) &  - & $\geq 2.5\times 10^4$ & - & - & 50 & $10^5$\\
\hline
\end{tabular}
\end{table*}

In order to take into account the expected non-LTE effects, we analysed
the data by means of a non-LTE LVG (Large Velocity Gradient)
code. The general details of the code are described in \citet{2002A&A...383..603C}.
We included the molecular data and the collisional coefficients for the
SO, SO$_2$ and CS molecules, by taking the former from the JPL Catalogue
\citep[\it{http://spec.jpl.nasa.gov/;}][]{JPL} and the latter from 
\citet{1994ApJ...434..188G},  \citet{1995ApJS..100..213G} and 
\citet{1992ApJ...399..114T} respectively. Briefly, the used non-LTE LVG code
computes the rotational line fluxes solving simultaneously the statistical
equilibrium of the levels and the radiative transfer of the emitted photons. The
approximation of the LVG consists of assuming that the photons emitted at any
point of the medium either are absorbed in the immediate vicinity or they escape
(this is the so-called photon escape formalism), because the different regions
are moving at high velocities. In practice, the photons cannot be
absorbed if their frequency -shifted by the Doppler effect- is outside the
thermal width centered on the rest frequency of the transition. The LVG
approximation assumes homogeneous and isothermal semi-infinite slab. Although
this approximation is strictly correct in the presence of large velocity gradient,
and the absence of temperature and density gradients, it is very often used to give
a first estimate of the average gas density and temperature of the observed
medium. In this work, we will use the non-LTE code in this way. More
sophisticated models would be required for a more sophisticated analysis.
However, given the quality of the obtained data, a greater sophistication of the
analysis would be unjustified.

The code has three free parameters to be constrained by comparing the
model predictions with observed data: the gas density and temperature 
and the column density of the species. Implicitly, we are assuming that the
emission fills the beam (24$''$) and the observed linewidth, namely 4 and 
5~km~s$^{-1}$ for HVC of SO and  SO$_2$ and 1~km~s$^{-1}$ for LVC of both 
molecules.
We ran several cases varying the SO and SO$_2$ column densities from 
$10^{13}$ to $10^{17}$ cm$^{-2}$, the density between
$10^4$ and $10^7$ cm$^{-3}$ and the temperature between 10 and 200 K.
We then searched for the minimum $\chi^2$ in this 3-D space, where
the $\chi^2$ is defined as usual :
$$\chi^{2}= \frac{1}{N-2} \sum_{1}^{N} 
\frac{(Observations-Model)^{2}}{\sigma^{2}}$$

We computed the SO and SO$_2$ column densities, as well as
the temperature and density of the gas in five and  four positions of the 
west and east lobes of the outflow  respectively (Tables~\ref{colchi} and 
\ref{phys_chi}). 
These positions include the E, W1 and W2 regions discussed in previous 
sections. 

Taking the species column densities with the lowest absolute $\chi^2$
(Table~\ref{colchi}), we constrained the gas density and temperature 
by plotting the $\chi^2$  contour at 2$\sigma$ (95.4\% in confidence) as 
a function
of these two parameters (see Fig.~\ref{chisq}).
In general, a family of temperature and density can reproduce the observed 
fluxes, so that they are only approximatively constrained.
Table~\ref{phys_chi} reports the kind of constraints we obtained at each 
studied position.

Having available only two transitions for CS we could not carry out 
a similar method to constrain the gas density and temperature and the CS 
column density
simultaneously. We assumed instead the average density and temperature as 
suggested by 
the SO observations and reported in Table~\ref{phys_chi} (see the detailed 
discussion in section 5.3), and computed
the CS column density consequently using the LVG model. The results are 
reported in Table~\ref{colchi}. Note that CS having the same density and
temperature as SO implies that the two species are also chemically related. This
is a major assumption, which is partially supported by the theoretical
predictions that, in dense and warm gas, CS is mainly formed by C + SO 
\citep[thus SO
and CS are chemically linked;][]{1997ApJ...481..396C,2004A&A...422..159W}.
However, even our observations cast some doubt on the robustness of this
assumption, for the SO and CS emissions in the maps are slightly displaced. Not
having other possibilities, we will adopt this assumption, but in the following
we will keep in mind the limits of the assumption used.

\subsection{Modeling results}\label{Mod_resul}

{\it Gas temperature and density}

The non-LTE analysis gives an indication of the gas density and temperature
along the outflow and in the LVC and HVC components 
(Fig.~\ref{chisq} and
Table~\ref{phys_chi}), within the limits discussed in the previous section and 
summarized
again at the end of this section.
In the following, we discuss the two velocity components separately.\\

\noindent
{\it i) LVC component}\\
Although it is difficult to constraint exactly (with the presented 
observations)
the gas density and 
temperature in this component, the plot in Fig.~\ref{chisq} suggests densities 
between $2\times 10^4$ and $2\times 10^5$~cm$^{-3}$
and temperatures somewhat higher than 15~K.
Complementary CO J=2-1 line observations of the IRAS2 outflow indicate 
brightness temperatures of 
20-22~K, which implies a gas kinetic temperature of about 25~K (Wakelam  2005
in prep). This is 
consistent with 
\citet{1996A&A...306..935W} who derived a similar temperature (about 20 K)
and a similar density (about $10^5$~cm$^{-3}$) for the ambient gas by using 
$^{13}$CO and C$^{18}$O observations with a larger beamsize (2.5 arcmin).
Thus, for the modelling of the chemistry (next section), we will assume 
a density of $5\times 10^4$~cm$^{-3}$ and a temperature of 20~K for this 
component. 

\noindent
{\it ii) HVC component}\\
Here, the temperatures and densities are higher than in the LVC component and 
different in the two lobes of the outflow. 
Indeed, the gas temperature and density in the two lobes (East and West), and 
as derived from SO and SO$_2$ respectively, present a slightly different 
behaviour.
The temperature and density derived from SO$_2$ in the East lobe suggest
a gradient along the outflow. The two positions of maximum SO$_2$ emission 
at $(60\arcsec,-12\arcsec)$ and $(72\arcsec,-12\arcsec)$ display the 
highest  kinetic temperatures, whereas the other points, at the edge 
of the emission peak, display higher densities and lower temperatures.
The gas probed by the SO$_2$ emission seems therefore warmed at the
peak position, and compressed around it.
A similar behavior may also be present in the gas probed by the SO 
transitions but it is less evident.
Remarkably, Fig.~\ref{chisq} shows that the T-n curves of the SO transitions 
do not overlap the curves of the SO$_2$ transitions.
The SO data seem to probe a gas less dense than the gas probed by the SO$_2$
transitions, suggesting a spatial differentiation in the formation of 
these two molecules.
Indeed, the map of the emission extent of the HVC component
suggests that the SO emission is more extended than the SO$_2$.
However, the data are not sufficient to better quantify this, so that in the
following we will assume in our chemical modeling that both molecules 
originate
in the same gas, where the density is around $1\times10^5$ cm$^{-3}$ and
the temperature is about 50 K, the median values between the SO and SO$_2$
derivations. The same applies for CS, which is also assumed to originate in
the same gas the SO and SO$_2$. As previously mentioned (\S
5.2), this is likely a rough approximation, but the available data do not allow
a better refinement of the treatment.  \\

\noindent
{\it iii) Column densities and abundance ratios}\\
As anticipated, the column densities computed with the non-LTE LVG method 
(Table~\ref{colchi}) are very similar to the ones derived by the rotational 
diagrams. 
They differ by less than 30\% with the exception of the SO$_2$  column
density in W2(HVC) where the difference reaches 50\%. 
The uncertainty on this latter value is higher because two transitions of 
SO$_2$ were not detected in W2. 
Note that the SO and SO$_2$ transitions are predicted
to be optically thin at the derived column densities. 

Using the derived column densities, we estimated the abundance ratios
reported in Table~\ref{colchi} and
shown as function of the distance from the protostar in Fig.~\ref{cs_so_so2_d}.
The absolute abundances were derived from the CO column density, assuming 
a standard CO to $\rm H_2$ abundance $[CO]= 10^{-4}$, following
\citet{1996A&A...306..935W}. We adopted the 
value estimated by \citet{1996A&A...306..935W} for the LVC~: 
$8\times 10^{22}$~cm$^{-2}$. 
For the HVC, we took the CO column densities obtained from complementary
observations of the CO J=2-1 line for our study of the SiO emission in the 
IRAS2 protostellar outflow (Wakelam 2005, in prep). The CO column density 
was estimated 
 assuming the levels to be populated according to LTE and the transition to 
be optically thin. This assumption is motivated by the fact that the CO profiles
are similar to the CS 2-1 ones with a line width of 10-20~km~s$^{-1}$. However,
only $^{13}$CO 2-1 observations could confirm this hypothesis and the
real H$_2$ column density could be higher than the one derived here. 
We adopted a kinetic temperature of 50~K for the CO gas. 
The values in the HVC are close to $\rm 10^{17}cm ^{-2}$; they 
are summarized in Table~4 and are in agreement with the 
values obtained by Bergin et al. (2003), averaged over a region of 4 arcmin 
size.

We derive the following average absolute abundances for SO, SO$_2$ and CS 
respectively~: $9\times 10^{-10}$, $10^{-10}$ and $4\times 10^{-10}$ in the 
LVC. We obtain much higher values in the HVC~: 
$2\times 10^{-7}$, $4\times 10^{-8}$ and $4\times 10^{-8}$ in the 
West lobe, and $3\times 10^{-7}$, $7\times 10^{-8}$ and $2\times 10^{-8}$ 
in the East lobe. Unfortunatly, we cannot perform an accurate analysis at each 
 point because of the lack of observations. Therefore, we will 
use the abundance ratios for the subsequent modelling.

Remarkably, the derived abundance ratios are different in the LVC and HVC
components, which, therefore, differ not only in the kinematics and
gas temperature and density, but also in the chemical composition.
The CS/SO ratio is 0.3-0.6 in the LVC component, with the ratio steadily 
decreasing going outward (across the outflow).
On the contrary, the CS/SO ratio in the HVC component is much more variable: 
it is between 0.03 and 0.1 in the East lobe, and 0.1-0.3 in the West lobe, 
i.e. up to 6 times larger in the West lobe.
Contrary to the LVC component, it increases going outward in both lobes, 
with the exception of the farthest point in the West lobe.
However, the largest difference between the LVC and HVC components is seen in 
the SO$_2$/SO ratio; less than 0.005 in the LVC component 
and about 0.2 in the HVC component, almost constant along the outflow.\\

{\it iv) Uncertainties}\\
One has to keep in mind the various approximations/assumptions
used when trying to interpret these results. 
First, in our analysis we assumed that the three molecules originate
in the same gas, which may not be the case.
Second, in practice  we assumed that the emission fills the beam, or that
the filling factor is the same for the three molecules.
If the lines are optically thin this would give values (gas density and 
temperature, as well as the species column density) that are correct, but if 
the lines are optically thick this may not be the case. In practice, we 
estimate that the computed abundance ratios are likely uncertain by about 
60\%. Higher spatial resolution observations are required for better estimates.

\section{Excitation conditions}\label{exc_cond}
\begin{figure}
\centering
\includegraphics[angle=90,width=0.6\linewidth]{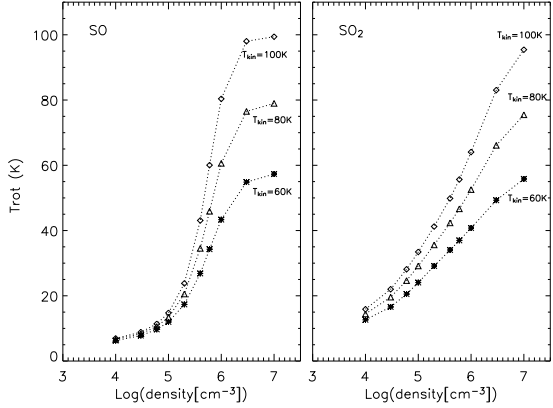}
\caption{SO (left panel) and SO$_{2}$ (right panel) rotational temperature 
as a function of density computed for several kinetic temperatures between
60 and 100~K. We assumed the SO and SO$_2$ column densities found in W1(HVC) (N$_{\rm SO} = 1.5\times 10^{14} cm^{-2}$ and N$_{\rm SO_2} = 3\times 10^{13} cm^{-2}$) using rotational diagrams. See text for details.}
\label{Trot}
\end{figure}
\begin{figure}
\centering
\includegraphics[width=0.7\linewidth]{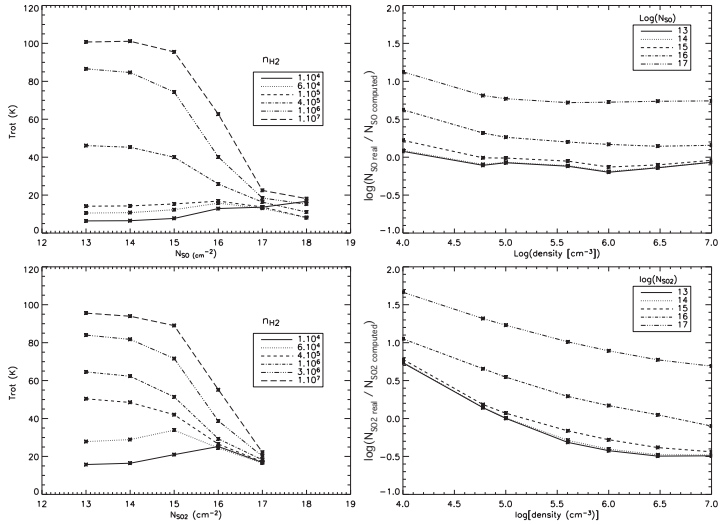}
\caption{Left pannels: SO and SO$_2$ rotational temperatures as a function of column densities computed for several H$_2$ densities between $10^4$ and $10^7$~cm$^{-3}$and a kinetic temperature of 100~K. Right pannels: Error on the SO and SO$_2$ column densities computed with rotational diagrams as a function of H$_2$ densities and for several column densities. }
\label{Trot_NSO}
\end{figure}
In this appendix, we present a study of the SO and SO$_2$ excitation in the shocked gas using a LVG model and following the \citet{1998A&A...335..266B} idea. In practice, we would like to understand what physical conditions can explain an underestimation of the rotational temperatures derived from the SO and SO$_2$ emission (section~\ref{LTE}). In addition, we studied the robustness of the column densities derived with the rotational diagrams if the basic hypothesis of the method is not verified. 

Using the SO and SO$_2$ fluxes computed by the LVG model, we reconstruct theoretical rotational diagrams for different column densities, kinetic temperatures and H$_2$ densities. From these rotational diagrams, we derived theoretical rotational temperatures and column densities. To compare the theoretical diagrams with the observed ones, we constructed them only using the LVG fluxes of the four observed transitions of SO and SO$_2$. The rotational diagram method is based on the assumptions of local thermal equilibrium (LTE) and of optically thin lines. We studied the effects of the first assumption by varying the H$_2$ density whereas the effects of the optical depth depend on the column density. 

{\bf Effects of the H$_2$ density:}\\
Assume first that the lines are optically thin ($\rm N_{SO}$ and $\rm N_{SO_2} < 10^{15}$~cm$^{-2}$). For a fixed column density, we computed the theoretical $\rm T_{rot}$ varying the kinetic temperature between 60 and 100~K and the H$_2$ density between $10^4$ and $10^7$~cm$^{-3}$ and show them on Fig.~\ref{Trot}. If the kinetic temperature of the gas is higher than 60~K, the computed $\rm T_{rot}$ can be as low as the observed ones (7-13~K for SO and 20-30~K for SO$_2$, see Table~\ref{resuldiag}) if the H$_2$ density is lower than $10^6$~cm$^{-3}$. The critical densities of the higher transitions for the two molecules are around $10^6$ and $10^7$~cm$^{-3}$ respectively. So if the gas density is lower than $10^6$~cm$^{-3}$, the LTE is not reached and the higher levels of the transitions are not populated, resulting in $\rm T_{rot} < T_{k}$. 

{\bf Effects of the optical depth:}\\
To study the effect of the optical depth, we varied the column density between $10^{13}$ and $10^{17}$~cm$^{-2}$ (see Fig.~\ref{Trot_NSO}). We assume LTE (i.e. $\rm n_{H_2} = 10^7$~cm$^{-3}$). The computed rotational temperatures decrease very quickly for SO and SO$_2$ column densities higher than $10^{15}$~cm$^{-2}$. If the kinetic temperature of the gas is higher than 100~K and the LTE reached, a SO and SO$_2$ column density as high as $10^{17}$~cm$^{-2}$ is needed to obtain $\rm T_{rot}$ as low as observed. Assuming an H$_2$ column density in the shocked gas of $5.8\times 10^{20}$~cm$^{-2}$ \citep{1998A&A...335..266B}, this gives an SO and SO$_2$ abundance of $10^{-4}$, which is not realistic. 

{\bf Effects on the computed column densities:}\\
In case of non-LTE and/or optically thick lines, the column densities computed with rotational diagrams can be wrong. To quantify the error, we computed the theoretical column densities for different real column densities, kinetic temperatures and H$_2$ densities. In Fig.~\ref{Trot_NSO}, we reported the ratios between the real column densities (used to run the LVG model) and the computed ones (derived from the theoretical diagrams) for all parameters. Whatever the physical conditions, SO is underestimated. The error is a factor of 1.5 if $\rm N_{SO real} \le 10^{15}$~cm$^{-2}$ whatever $\rm n_{H_2}$, a factor of 10 and 6 at low and high densities respectively if $\rm N_{SO real} = 10^{17}$~cm$^{-2}$. The error is much higher on $\rm N_{SO_2}$: for $\rm N_{SO_2 real} \le 10^{15}$~cm$^{-2}$, $\rm N_{SO_2 computed}$ is underestimated by a factor of 6 at low densities and overestimated by a factor of 3 at LTE. 

Thus, the most probable explanation for the low observed $\rm T_{rot}$ is a non-LTE effect because of H$_2$ densities lower than $10^6$~cm$^{-3}$. Considering the observed column densities arounf $10^{14}$~cm$^{-3}$ (Table~\ref{colchi}), the error on the computed column densities should not be high: less than a factor of 2 for SO and 6 for SO$_2$ depending on the H$_2$ density.

\end{document}